\documentclass[twocolumn,showpacs,preprintnumbers]{revtex4}
\usepackage{graphicx}
\usepackage{dcolumn}
\usepackage{bm}
\usepackage{epsfig}
\usepackage{amsmath}
\usepackage{subfigure}

\begin{document}

\title{Bidirectional and tunable single-photons multi-channel quantum router between microwave and optical light}
\author{Peng-Cheng Ma$^{1,2,3}$}
\author{Jian-Qi Zhang$^{2}$}
\email{changjianqi@gamail.com}
\author{Mang Feng$^{2}$}
\email{mangfeng@wipm.ac.cn}
\author{Zhi-Ming Zhang$^{1}$}
\email{zmzhang@scnu.edu.cn}
\address{$^1$Laboratory of Nanophotonic Functional Materials and Devices (SIPSE), and
Laboratory of Quantum Engineering and Quantum Materials, South China
Normal University, Guangzhou 510006, China\\
$^2$State Key Laboratory of Magnetic Resonance and Atomic and Molecular Physics, Wuhan Institute of Physics and Mathematics,
Chinese Academy of Sciences, Wuhan 430071, China
\\ $^3$School of Physics
and Electronic Electrical Engineering, Huaiyin Normal University,
Huaian 223300, China}

\begin{abstract}
Routing of photon play a key role in optical communication and
quantum networks. Although the quantum routing of signals has been
investigated in various  systems both in theory and experiment.
However, no current theory can route quantum signals between
microwave and optical light. Here, we propose an experimentally
accessible tunable multi-channel quantum routing proposal using
photon-phonon translation in a hybrid opto-electromechanical system.
It is the first demonstration that the single-photon of optical
frequency can be routed into three different output ports by
adjusting microwave power. More important, the two output signals
can be selected according to microwave power. Meanwhile, we also
demonstrate the vacuum and  thermal noise will be insignificant for
the optical performance of the single-photon router at temperature
of the order of 20 mK. Our proposal may have paved a new avenue
towards multi-channel router and quantum network.
\end{abstract}
 \maketitle

Quantum information science has been developed rapidly due to the
substitution of photons as signal carriers rather than the limited
electrons \cite{nature-453-1023}. Single photons are suitable
candidates as the carrier of quantum information due to the fact
that they propagate fast and interact rarely with the environment.
Meanwhile, a quantum single-photons router is challenging because
the interaction between individual photons is generally very weak.
Quantum router or quantum switch plays a key role in optical
communication  networks  and quantum information processing. It is
important for controlling the path of the quantum signal with fixed
Internet Protocol (IP) addresses, or quantum switch without fixed IP
addresses.

Designing a quantum router or  an optical switch operated at a
single photon level enables a selective quantum channel in quantum
information and quantum networks
\cite{prl-106-053901,prl-107-073601,pra-85-021801,nature-508-241,Science-341-768,
prl-111-193601, Nat.Photon-7-373, nat.phy.-3-807}, such as in
different systems, cavity QED system \cite{prl-102-083601}, circuit
QED system \cite{prl-107-073601}, optomechanical system
\cite{pra-85-021801}, a pure linear optical system
\cite{pra-83-043814,arXiv:1207.7265}, $\wedge$-type three-level
system \cite{prx-3-031013,prl-111-103604,pra-89-013806}. The essence
lying at the core is the realization of the strong coupling between
the photons and photons or photons and phonons
\cite{prl-108-093604,Nat.Photon-6-605,Nat.Photon-4-477,Nat.Photon-2-185},
but these methods require high-pump-laser powers due to the very
weak optical nonlinearity. To the best of  our knowledge, the
quantum router demonstrated in most experiments and theoretical
proposals has only one output terminal, except for only a few
theoretical method in Ref.
\cite{prx-3-031013,prl-111-103604,pra-89-013806, scirep-4-4820} and
the experiments in Ref.\cite{arXiv:1207.7265}. However, until now,
the above all of routers are applied only in optical light or only
in microwave separately.

On the other hand, with the technological advancements in the fields
of optical nanocavity and microwave circuit, it is now possible to
engineer interactions between optical and microwave using
photon-phonon translation. Both microwave and optical light have
been separately used to cool a nanomechanical-resonator (NR)  to its
quantum ground state of motion \cite{nature-475-359, nature-478-89}
and work in the strong coupling regime \cite{nature-471-204,
science-330-1520}. This same interaction enables the mechanical
resonator to serve as an information storage
medium\cite{nature-482-63, nature-495-210}, and opens up the
possibility of high-fidelity frequency conversion
\cite{prl-108-153603,prl-108-153604,nature-495-201,nat.phys.-10-321}.

In this letter, by combining the technologies of optomechanics and
electromechanics, we   simultaneously couple a NR to both a
microwave circuit and an optical toroidal  nanocavity.   We show it
is the first demonstation that the multi-channel quantum router can
be tunable between microwave and optical light in this hybrid
system. More important, the two output signals frequencies can be
selected according to the microwave power. The thermal noise could
be more critical in deteriorating the performance of the
single-photon router. Then, we also demonstrate the vacuum and
thermal noise can be insignificant for the optical performance of
the single-photon router at temperature of the order of 20 mK.

\emph{Model setup and  solutions}.----
\begin{figure}[tbp]
\centering
\includegraphics[width=8cm,height=5cm]{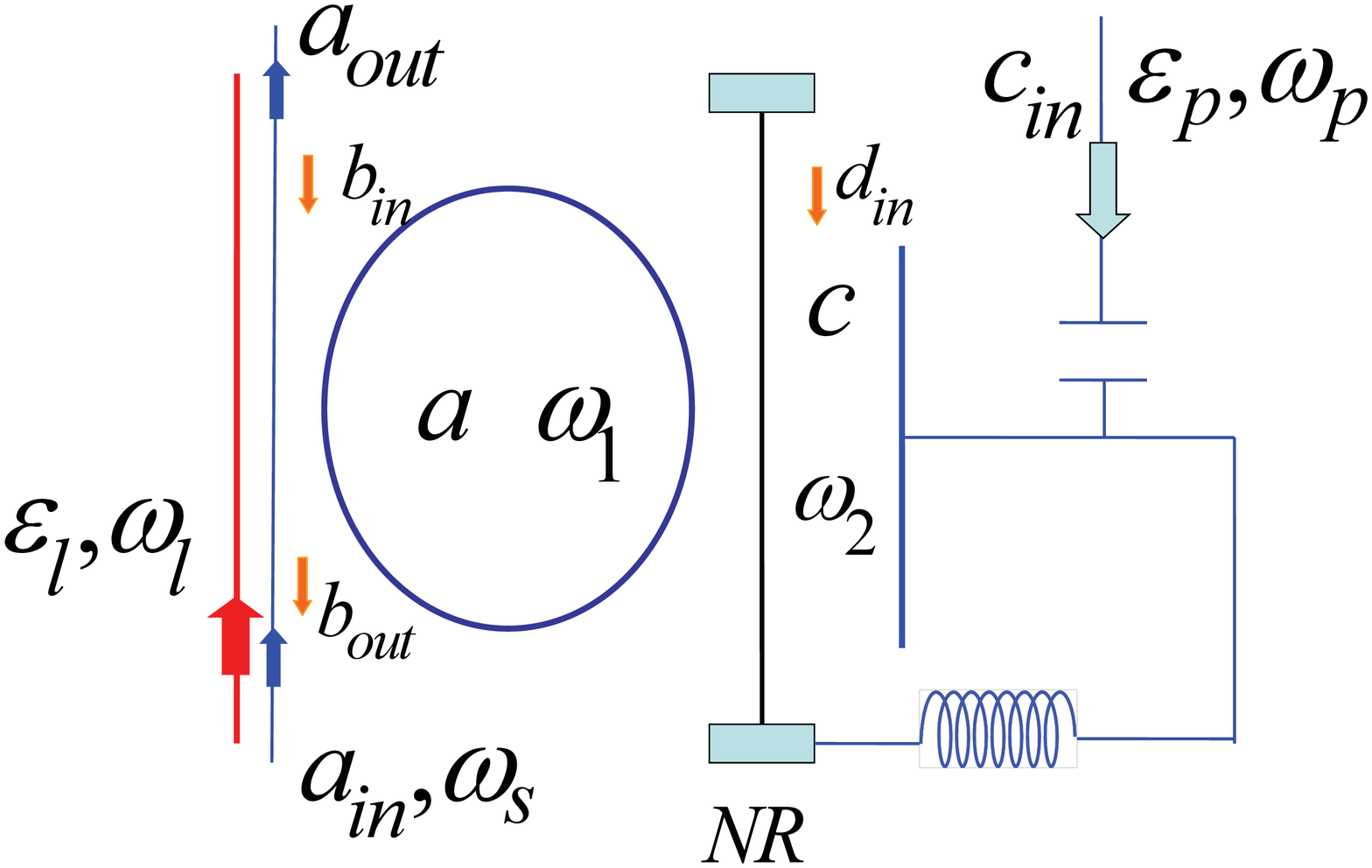}
\caption{(Color online) Schematic diagram of the hybrid system.
Which consist of microwave circuit and an optical toroidal
nanocavity are coupled to a common NR via capacitively coupling and
evanescent coupling, respectively. The optical toroidal nanocavity
with resonance frequency  $\omega_1$ is driven by  a strong pump
beam $\varepsilon_l$ with frequency  $\omega_l$ and a probe beam in
a single-photon Fock state with frequency $\omega_s$ simultaneously.
The microwave   cavity with resonance frequency $\omega_1$ is driven
by a strong pump beam $\varepsilon_p$ with frequency $\omega_p$. }
\end{figure}
The model for realizing tunable and bidirectional multi-channel
quantum router is sketched in Fig. 1, where a microwave circuit and
an optical toroidal nanocavity are coupled to a common NR  via
capacitively coupling \cite{nature-471-204} and evanescent coupling
\cite{nat.phy-5-909}, respectively. The optical toroidal nanocavity
with resonance frequency  $\omega_1$ is driven by  a strong pump
beam $\varepsilon_l$ with frequency  $\omega_l$ and a probe beam in
a single-photon Fock state with frequency $\omega_s$ simultaneously.
The microwave  circuit with resonance frequency  $\omega_2$ is only
driven by a strong pump beam $\varepsilon_p$ with frequency
$\omega_p$. In the rotating frame at the frequency $\omega_l$ and
$\omega_p$, the Hamiltonian of the hybrid system can be written as
\begin{eqnarray}
H=\hbar\Delta_aa^\dag a+\hbar\Delta_cc^\dag c
+\frac{p^{2}}{2m}+\frac{1}{2} m\omega _{m}^{2}q^{2}-\hbar g_1a^{\dag }aq\nonumber \\
+\hbar g_2c^{\dag }cq +i\hbar \varepsilon _{l}(a^{\dag }-a)+i\hbar
\varepsilon _{p}(c^{\dag }-c).
\end{eqnarray}
Here, $a \  (a^{\dag })$ and   $c\ (c^{\dag })$ are the annihilation
(creation) operator of optical toroidal nanocavity and microwave
circuit, respectively. $\Delta_a=\omega_1-\omega_l$,
$\Delta_c=\omega_2-\omega_p$ are the corresponding cavity-pump field
detunings. $p$ and $q$ are the momentum and the position operator of
the NR, respectively. The NR with frequency $\omega_{m}$ and
effective mess $m$. $g_1$($g_2$) is the single-photon coupling rate
between the mechanical mode and the optical (microwave) mode
\cite{nature-452-72}. The last two  terms in Eq.(1) describe the
interaction between the cavity field (optical and microwave) with
the input fields. The pump field strength $\varepsilon_{l}$
($\varepsilon_{p}$)  depends on the power $\wp_l$($\wp_p$) of
coupling field,
$\varepsilon_{l}=\sqrt{2\kappa_1\wp_{l}/\omega_{l}}$,
($\varepsilon_{p}=\sqrt{2\kappa_2\wp_{l}/\omega_{p}}$) with
$\kappa_1$ ($\kappa_2$)is the optical toroidal nanocavity (microwave
circuit) decay rate.

 Note that the   NR coupled to the thermal surrounding at the
temperature $T$, which results in the mechanical damping rare
$\gamma_m$, and thermal noise force $\xi$  with frequency-domain
correlation \cite{pra-85-021801},
\begin{eqnarray}
\langle \xi(\omega)\xi(\Omega)\rangle=2\pi\hbar\gamma_m m \omega[1+
\coth(\frac{\hbar\omega}{2\kappa_BT})]\delta(\omega+\Omega),
\end{eqnarray}
where   $\kappa_B$ is the Boltzmann constant. In addition, the
cavity field $a$ is coupled to the input quantum fields $a_{in}$ and
$b_{in}$. If there are no photons incident from the other direction,
then $b_{in}$ would be the vacuum field. Let $2\kappa_1$ be the
decay rate at which photons leak out from   the optical toroidal
nanocavity.
  The output fields can be written as
\begin{eqnarray}
x_{out}(\omega)=\sqrt{2\kappa_1}a(\omega)-x_{in}(\omega),\ \  x=a, \
b.\label{3}
\end{eqnarray}
These couplings are included in the standard way by writing quantum
Langevin equations for the cavity field operators with the
communtation relations $[a,a^{\dag }]=1$, $[c,c^{\dag }]=1$,
$[p,q]=i\hbar$. Putting together all the quantum fields, thermal
fluctuations, and the Heisenberg equations from the Hamiltonian (1),
we can obtain the working quantum Langevin equations:
\begin{eqnarray}
&&\dot{a}=-[2\kappa_1 +i(\Delta _a+g_1q)]a+\varepsilon
_{l}+\sqrt{2\kappa_1}a_{in}+\sqrt{2\kappa_1}b_{in}, \notag \\
&&\dot{c}=-[2\kappa_2 +i(\Delta _c-g_2q)]c+\varepsilon
_{p}+\sqrt{2\kappa_2}c_{in}+\sqrt{2\kappa_2}d_{in}, \notag \\
&&\dot{q}=\frac{p}{m},\ \ \  \notag   \\
&&\dot{p}=-m\omega _{m}^{2}q-\hbar g_1 a^{\dag }a+\hbar g_2 c^{\dag
}c-\gamma _{m}p+\xi,\label{4}
\end{eqnarray}
The quantum Langevin equations (\ref{4}) can be solved after all
operator are linearized as its steady-state mean value and a small
fluctuation:
\begin{eqnarray}
 p=p_{s}+\delta p,\  q=q_{s}+\delta q, \ a=a_s +\delta a, \ c=c_s +\delta c, \label{5}
\end{eqnarray}
where $\delta p$, $\delta q$,  $\delta a$, $\delta c$ being the
small fluctuations around the corresponding steady values. After
substituting Eq.(\ref{5}) into  Eq.(\ref{4}), ignoring the
second-order small terms, and introducing the Fourier transforms
$f(t)=\frac{1}{2\pi}\int_{-\infty}^{+\infty}f(\omega)e^{-i\omega
t}d\omega$,
$f^+(t)=\frac{1}{2\pi}\int_{-\infty}^{+\infty}f^+(-\omega)e^{-i\omega
t}d\omega$. We can get the steady values
\begin{eqnarray}
&&p_{s}=0, \ \   q_{s}=\frac{\hbar g_2|c_{s}|^{2}-\hbar
g_1|a_{s}|^2}{m\omega_m^2},  \notag \\
&&a_{s}=\frac{\varepsilon _{l}}{2\kappa_1+ i\Delta_1 },\ \
c_{s}=\frac{\varepsilon _{p}}{2\kappa_2+ i\Delta_2},
\end{eqnarray}
with $\Delta_1 =\Delta _a+g_1q_s$,  $\Delta_2 =\Delta _c-g_2q_s$ and
the solution of $\delta a$ \cite{pra-83-043826},
\begin{eqnarray}
\delta a
&&=E_1(\omega)a_{in}(\omega)+F_1(\omega)a^+_{in}(-\omega)+E_1(\omega)b_{in}(\omega) \nonumber \\
&&+F_1(\omega)b^+_{in}(-\omega)+E_2(\omega)c_{in}(\omega)+F_2(\omega)c^+_{in}(-\omega) \nonumber \\
&&+E_2(\omega)d_{in}(\omega)
+F_2(\omega)d^+_{in}(-\omega)+V(\omega)\xi(\omega),\label{7}
\end{eqnarray}
in which,
\begin{eqnarray}
&&E_1(\omega)=
\frac{-i\hbar\sqrt{2\kappa_1}}{d(\omega)}(|a_{s}|^2g_1^2A_2B_2
\nonumber
\\&&\ \ \ \ \ \ \ \ \     +2|c_{s}|^2g_2^2A_1\triangle_2 +m N A_1A_2B_2),
\nonumber \\
&&F_1(\omega)=\frac{-i\hbar\sqrt{2\kappa_1}|a_{s}|^2g_1^2A_2B_2}{d(\omega)},\nonumber \\
&&E_2(\omega)=\frac{-i\hbar\sqrt{2\kappa_2}a_{s}c_{s}g_1g_2A_1A_2}{d(\omega)},\nonumber \\
&&F_2(\omega)=\frac{i\hbar\sqrt{2\kappa_2}a_{s}c_{s}g_1g_2A_1B_2}{d(\omega)},\nonumber \\
&&V(\omega)=\frac{a_{s}g_1A_1A_2B_2}{d(\omega)}, \label{8}
\end{eqnarray}
with
\begin{eqnarray}
&&d(\omega)=2\hbar|a_{s}|^2g_1^2 \Delta_1A_2B_2+2\hbar|c_{s}|^2g_2^2
\Delta_2A_1B_1\nonumber \\
&&\ \ \ \ \ \ \  +m N A_1B_1A_2B_2,\nonumber \\
&&A_1=\Delta_1+\omega+2i\kappa_1,\ B_1=\Delta_1-\omega-2i\kappa_1,\nonumber \\
&&A_2=\Delta_2+\omega+2i\kappa_2,\ B_2=\Delta_2-\omega-2i\kappa_2,\nonumber \\
&&N=\omega^2+i\omega\gamma_m-\omega_m^2.\label{9}
\end{eqnarray}

\begin{figure}
\begin{minipage}[b]{0.5 \textwidth}
\includegraphics[width=0.5 \textwidth]{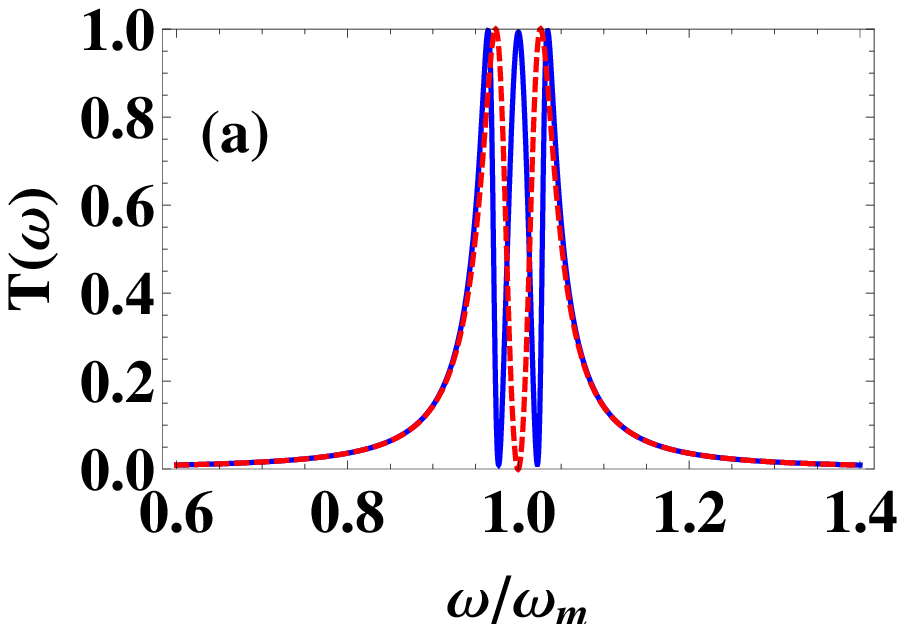}\includegraphics[width=0.5 \textwidth]{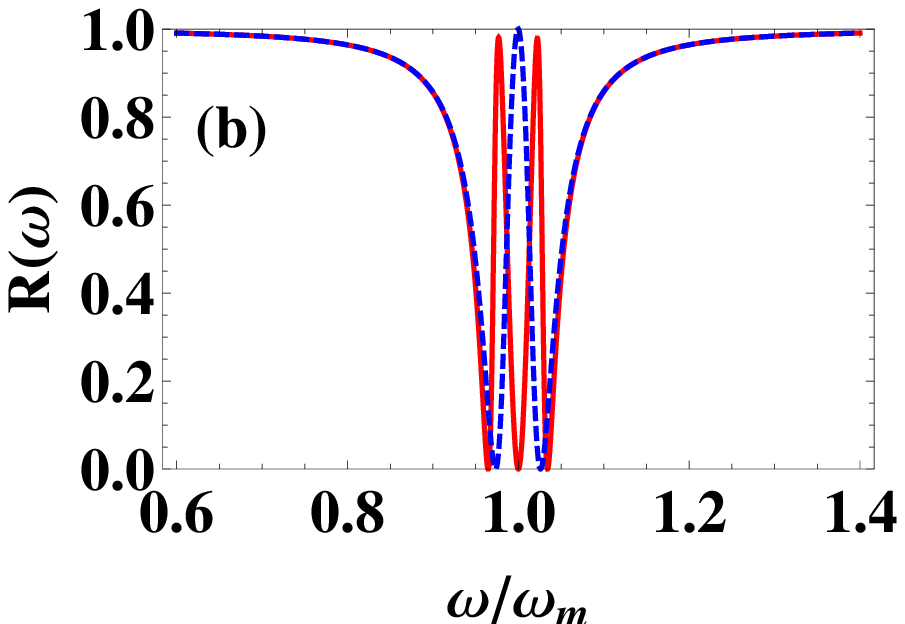}
\end{minipage}
\caption{(Color online) (a)  The transmission spectrum $T(\omega)$
and (b) the reflection spectrum $R(\omega)$  of the single photon as
a function of normalized frequency $\omega/\omega_m$ when microwave
pump filed is turned off (red dashed line) and turned on (blue solid
line). $\omega_p=2\pi\times7.1$GHz,  $m=48\ $ng,
$\omega_m=2\pi\times10.56$MHz, $\gamma_m=2\pi\times32$Hz,
$\kappa_1=4\pi\times100$KHz,
$\kappa_2=2\pi\times1$KHz,$\wp_l=2\times65\ \mu$W, $\wp_p=6\times50\
$nW \cite{nature-452-72, nature-471-204,nat.phy-5-909}. }
\end{figure}

Defining the spectrum  of the field via $\langle
a^+(-\Omega)a(\omega)\rangle=2\pi S_a(\omega)\delta(\omega+\Omega)$,
$\langle a(\omega)a^+(-\Omega)\rangle=2\pi
[S_a(\omega)+1]\delta(\omega+\Omega)$. The incoming vacuum field
$b_{in}$ and  $d_{in}$ are characterized by $\langle
\tau(\omega)\tau^+(-\Omega)\rangle=2\pi\delta(\omega+\Omega)$
$(\tau=b,d)$ with $S_{bin}=S_{din}(\omega)=0$. From Eq. (\ref{3})
and Eq. (\ref{7}), we find that  the spectrum of the output  fields
has the form,
\begin{eqnarray}
&&S_{aout}(\omega)=R(\omega)S_{ain} +S^{(T)}(\omega)+S^{(V)}(\omega), \nonumber \\
&&S_{bout}(\omega)=T(\omega)S_{ain} +S^{(T)}(\omega)+S^{(V)}(\omega),\nonumber \\
\label{10}
\end{eqnarray}
where
\begin{eqnarray}
&&R(\omega)=|\sqrt{2\kappa_1}E_1(\omega)-1|^2, \
T(\omega)=|\sqrt{2\kappa_1}E_1(\omega)|^2,  \nonumber \\
&&S^{(T)}(\omega)=2\kappa_1 |V(\omega)|^2 \hbar\gamma_m
m(-\omega)[1+
\coth(\frac{-\hbar\omega}{2\kappa_BT})], \nonumber \\
&&S^{(V)}(\omega)=4\kappa_1 |F_1(\omega)|^2. \label{11}
\end{eqnarray}

\emph{Single-photons multi-channel quantum router between microwave
and optical light}. ----  In Eq.(\ref{10}), $R(\omega)$ and
$T(\omega)$ are the contributions arising from the presence of a
single photon in the input field. $S^{(v)}(\omega)$ is the
contribution from incoming vacuum field. The $S^{T}(\omega)$ is the
contributions from the fluctuation of   the NR, respectively.
Eq.(\ref{10}) shows that even if there were no incoming photon, the
output signals is generated via quantum and thermal noises. For the
purpose of achieving a single-photons multi-channel quantum router,
the key quantities are $R(\omega)$ and $T(\omega)$. Further, we also
demonstrate the performance of the single-photon quantum router
should not be deteriorated by the quantum and thermal noises terms
$S^{(V)}(\omega)$ and $S^{T}(\omega)$.

To demonstrate the routing functions of the hybrid
opto-electromechanics system, we first investigate the reflection
$R(\omega)$ and transmission spectrum $T(\omega)$. For illustration
of the numerical results, we choose the realistically reasonable
parameters from the recent experiment \cite{nature-452-72,
nature-471-204,nat.phy-5-909}. $\omega_p=2\pi\times7.1$GHz,  $m=48\
$ng, $\omega_m=2\pi\times10.56$MHz, $\gamma_m=2\pi\times32$Hz,
$\kappa_1=2\pi\times100$KHz,
$\kappa_2=2\pi\times1$KHz,$\wp_l=2\times65\ \mu$W, $\wp_p=6\times50\
$nW. We also apply the following conditions \cite{pra-81-041803,
science-330-1520}, (i) $\Delta_1=\Delta_1\simeq\omega_m$ and (ii)
$\omega_m\gg\kappa_1$. The first condition means that the optical
cavity is driven by a red-detuned laser field which is on resonance
with the optomechanical anti-Stokes sideband. The second condition
is the well-known resolved sideband condition, which ensures the
normal mode splitting to be distinguished \cite{science-330-1520}.

\begin{figure}[tbp]
\centering
\includegraphics[width=4.4cm,height=3.1cm]{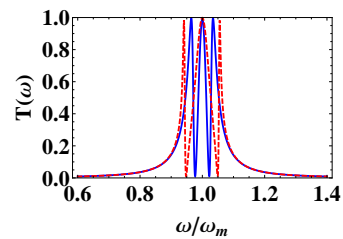}
\caption{(Color online) The transmission spectrum $T_1(\omega)$ of
the single photon as a function of normalized frequency
$\omega/\omega_m$  with different microwave pump filed power:
$\wp_p=6\times50\ $nW(blue solid line) and $\wp_p=30\times50\
$nW(red dashed line). Other parameters take the same values as in
Fig. 2.}
\end{figure}

The resulting spectra are shown in Fig 2. In the absence of the
microwave pump field (\emph{i.e.} $\wp_p=0$). One can observes an
inverted EIT and a standard EIT in the reflection and transmission
spectra of a single photon. Note the $R(\omega_m)\approx1$ and
$T(\omega)\approx0$. So the single photon is complete reflected at
frequency $\omega_m$. However, in the presence of the microwave pump
field, the situation is completely different $R(\omega_m)\approx0$
and $T(\omega_m)\approx1$. More important, the reflection and
transmission spectra of a single photon exhibit other two inverted
dips and two normal dips at $\omega=\omega_m+\omega_0$ and
$\omega=\omega_m-\omega_0$, here $\omega_0$ is the small deviation
from the central frequency $\omega_m$ which  depend on the microwave
pump field power. We can find $R(\omega_m\pm\omega_0)\approx1$ and
$T(\omega_m\pm\omega_0)\approx0$. That is to say, the single photon
is completely transmitted  at frequency $\omega=\omega_m$,
meanwhile, the single photons is completely reflected  to the other
two output ports at different frequencies $\omega_m+\omega_0$ and
$\omega_m-\omega_0$. The physical effect can be explained by
optomechanically induced transparency (OMIT)
\cite{science-330-1520,nature-452-69} which originates from the
radiation pressure coupling an optical mode to a mechanical mode.
The OMIT depends on quantum interference and it is sensitive to
phase disturbances. The coupling between microwave and the common NR
breaks down the symmetry of the OMIT interference, then the single
OMIT  window is split into two transparency windows
\cite{nat.phys.-8-891,ma2014}.

Now, we can describe the working process of the multi-output quantum
router between microwave and optical. When we turn off the
  microwave pump field, the single photons is complete reflected
at frequency $\omega=\omega_m$ (\emph{i.e.} $R(\omega_m)\approx1$,
$T(\omega_m)\approx0$). However, when we turn on the microwave pump
field, the single photons is complete transmitted  at frequency
$\omega=\omega_m$ ( \emph{i.e.} $R(\omega_m)\approx0$,
$T(\omega_m)\approx1$), at the same time, three are completely
reflected to the other two outputs at different frequencies
$\omega=\omega_m+\omega_0$ and $\omega=\omega_m-\omega_0$
(\emph{i.e.} $R(\omega_m\pm\omega_0)\approx1$ and
$T(\omega_m\pm\omega_0)\approx0$). Fig. 3 describe the transmission
spectrum $T(\omega)$ with  the different microwave pump field power.
From Fig. 3, we find the different output frequencies can be
selected by adjusting the microwave pump field power, then it can be
controlled the quantum signals to different IP address in quantum
networks. Similarly, we can use the same principle to route the
microwave signals by tuning optical light power, here, we do not
describe it one by one.

\begin{figure}
\begin{minipage}[b]{0.5 \textwidth}
\includegraphics[width=0.5 \textwidth]{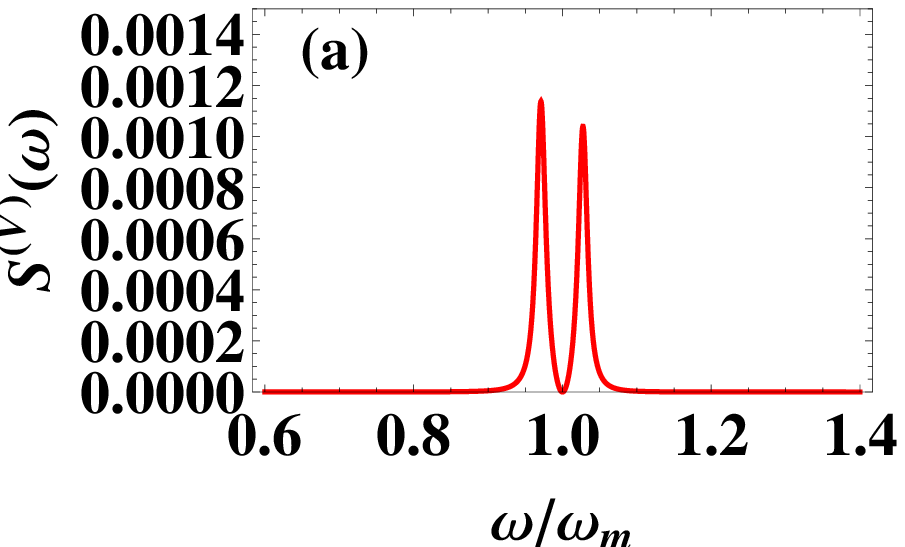}\includegraphics[width=0.5 \textwidth]{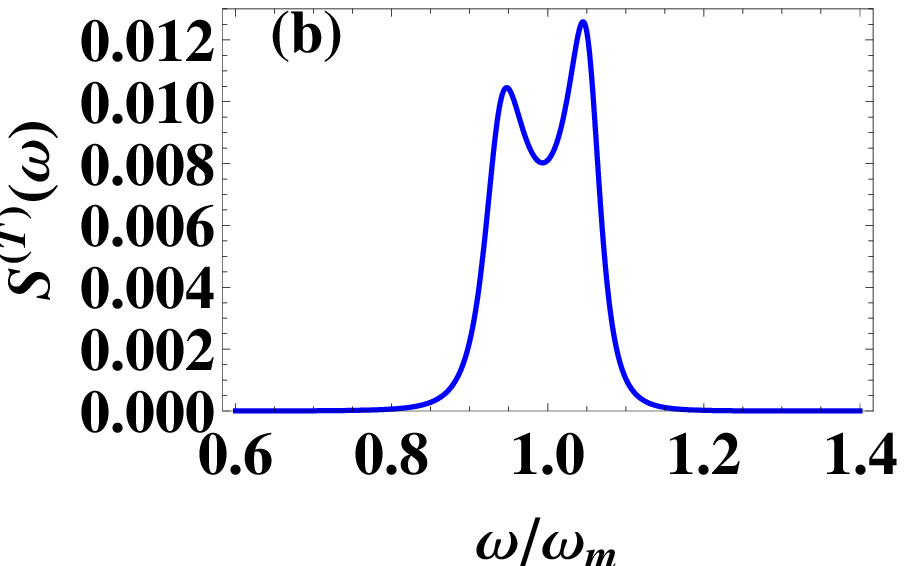}
\end{minipage}
\caption{ (a) The vacuum noise spectrum $S^{(V)}(\omega)$ as a
function of normalized frequency $\omega/\omega_m$.  (b) The thermal
noise spectrum $S^{(T)}(\omega)$ as a function of normalized
frequency $\omega/\omega_m$.  $T=20$mK, other parameters take the
same values as in Fig. 2.}
\end{figure}

Moreover, we discuss the effects of the quantum and thermal noise on
the reflection and transmission spectrum of a single-photon. From
Fig.4, the contribution of the vacuum noise maximum is about
$0.13\%$ and is thus insignificant. The thermal noise could be more
critical in deteriorating the performance of the single-photon
router. Clearly to beat the effects of thermal noise, the number of
photons in the probe pulse has to be much bigger than the thermal
noise photons. However, if we work with NR temperatures like 20 mK,
then the thermal noise term is insignificant as shown in Fig. 4(b).

\emph{In conclusion.}---- In this letter, we have proposed an
experimentally accessible a bidirectional and tunable single-photons
multi-channel quantum router between microwave and optical light
based on the hybrid opto-electromechanical system. The system
consist of a microwave circuit and an optical toroidal nanocavity
are coupled to a common NR. It is the first demonstration that the
single-photon of optical frequency can be routed into three
different output ports by adjusting microwave power. More important,
the two output optical signals can be selected according to
microwave power, then it can be controlled the quantum signals to
different IP address in quantum networks. Meanwhile, we also
demonstrate the vacuum and thermal noise will be insignificant for
the optical performance of the single-photon router at temperature
of the order of 20 mK. Our proposal may have paved a new avenue
towards multi-channel router and quantum network.

\section*{ACKNOWLEDGMENTS}
 PCM thanks Lei-Lei Yan  for their helps in
the numerical simulation.  This work was supported by the Major
Research Plan of the NSFC (Grant No.91121023), the NSFC (Grants No.
61378012, No. 60978009, No. 11274352 and No. 11304366), the
SRFDPHEC(Grant No.20124407110009), the "973"Program (Grant Nos.
2011CBA00200, 2012CB922102 and 2013CB921804), the PCSIRT (Grant
No.IRT1243). China Postdoctoral Science Foundation (Grant No.
2013M531771 and No. 2014T70760). Natural Science Fund for colleges
and universities in Jiangsu Province (Grant No.12KJD140002).

\end{document}